\documentclass[10pt,letterpaper,twocolumn]{article}
\usepackage{graphicx}
\usepackage{amssymb}
\usepackage{float}

\newcommand{\beq}{\begin{equation}}
\newcommand{\eeq}{\end{equation}}
\begin{document}

\title{Scale Free Networks of Earthquakes and Aftershocks}
\author{Marco Baiesi$^{*,\dag}$ and Maya Paczuski$^\dag$} 

\maketitle

$^*$INFM-Dipartimento di Fisica,
Universit\`a di Padova, I-35131 Padova, Italy.
$\dag$Mathematical Physics, Department of Mathematics,  Imperial College London,
London UK SW7 2BZ.

\begin{abstract}
{ We propose a new metric to quantify the correlation between any two
earthquakes. The metric consists of a product involving the time
interval and spatial distance between two events, as well as the
magnitude of the first one.  According to this metric, events
typically are strongly correlated to only one or a few preceding
ones. Thus a classification of events as foreshocks, main shocks or
aftershocks emerges automatically without imposing predefined
space-time windows.  To construct a network, each earthquake receives
an incoming link from its most correlated predecessor.  The number of
aftershocks for any event, identified by its outgoing links, is found
to be scale free with exponent $\gamma = 2.0(1)$.  The original Omori
law with $p=1$ emerges as a robust feature of seismicity, holding up to
years even for aftershock sequences initiated by intermediate magnitude
events.  The measured fat-tailed distribution of distances between
earthquakes and their aftershocks suggests that aftershock collection
with fixed space windows is not appropriate.  }
\end{abstract}

%%%%%%

\section{Introduction}
Earthquakes exhibit complex correlations in space, time, as well as
magnitude~\cite{turcotte_book,scholz_book,kagan,gutenberg-richter41,omori94b,bak02:_unified}.
Sequences of earthquakes often appear related to main shocks of large
magnitude, which are followed in time by nearby smaller events.
Sometimes, the main shock is also preceded by a few intermediate or
smaller precursor events. Earthquakes can also cluster as swarms,
where the seismic activity is not distinctly associated with a main
event.  Human observation tends toward labeling these events
depending on their relative magnitude and their position in the
space-time sequence: foreshocks, main shocks and aftershocks,
respectively.  However, in identifying aftershocks, it is necessary to
distinguish them from what is called background seismicity, and to identify
their main shock.  Although
an observation by eye of the evolving seismic situation can support
a classification, a precise label for each event may be
intrinsically impossible.

\begin{figure}[!t]
\begin{center}
\includegraphics[width=35mm,angle=-90]{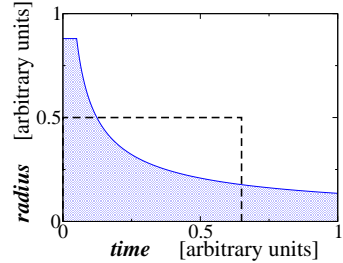}
\end{center}
\caption{
\label{fig:window}
Schematic examples of space-time windows used to collect aftershocks:
the usual rectangular or convex window (dashed line) 
and our hyperbolic, concave window (shaded region).
}
\end{figure}

In the most popular approach,  aftershocks are collected by
counting all events within a fixed space-time 
window~\cite{gardner74:_after_window,keilis80:_after_window,knopoff82:_after_window,knopoff00:_after_window} 
following a pre-assigned main event (see Fig.~\ref{fig:window}). However, this
method does 
not define the  probability that an event thereby collected is actually
correlated to the main event under consideration. Maybe more
importantly, one does not know whether the predefined space-time
windows are too large or too small for minimizing errors in the
procedure.  A more subtle issue is to
define aftershocks of aftershocks.  If an aftershock can have
more than one preceding large event, which of these should be
regarded as the most important or correlated one?

%FIG %%%
%FIG %%%

A quantitative metric of the correlation between any two earthquakes, or the extent to
which one can be considered an aftershock of
another, may be crucial for solving these problems, and
for developing a better understanding of seismicity.  A reliable metric should
include known statistical properties.  
One is the  Gutenberg-Richter (G-R) 
distribution~\cite{gutenberg-richter41}
for the number of earthquakes   of magnitude $m$ in a seismic region,
\beq P(m)   \sim 10^{-b\,m} \,,
\label{eq:G-R}
\eeq with $b$ usually $\approx 1$.  Another is the 
fractal appearance of earthquake
epicenters~\cite{turcotte_book,kagan,hirata89:_d_fract},
with fractal dimension $d_f$. Thus, the average number of earthquakes of
magnitude within an interval $\Delta m$ of $m$, occurring in an area of radius $r$
over a time interval $\tau$, is 
\beq 
\bar{n} = C\, \tau\, r^{d_f} \Delta m \, 10^{-b m}\,, 
\label{eq:n-mean}
\eeq 
where $C$ is a  constant depending on the overall seismicity in the region under consideration.

For any earthquake $j$ in the seismic region, looking backward in time, 
 how many earthquakes of magnitude within an interval $\Delta m$ of $m$
would be expected to have occurred within a time interval $t$, 
and within a distance $l$, of that specific event?   
In fact, an $n$ value can be defined
between any two events $i$ and $j$ occurring in the sequence at times
$T_i$ and $T_j$, with $T_i<T_j$.  If we take the magnitude $m_i$ of
the $i$-th event, the spatial distance $l=l_{ij}$ between the two
earthquake epicenters, and the time interval $t=t_{ij}=T_j-T_i$, the expected
number of events of magnitude within $\Delta m$ of $m_i$ occurring in the 
particular space-time domain bounded by events $j$ and $i$  is 
\beq 
n_{ij} \equiv C\, t\, l^{d_f} \Delta m \, 10^{-b m_i}\,.
\label{eq:n}
\eeq 
Note that the domain appearing in Eq.~\ref{eq:n} is selected by 
the particular history of the seismic activity in the region and not predefined by any observer. 
 
Of all the earthquakes preceding $j$, the most unlikely to
occur according to Eq.~\ref{eq:n} is earthquake $i^*$ such that
$n_{ij}$ is minimized when $i=i^*$.  However,
earthquake $i^*$ {\em actually}  occurred
relative to $j$, even though it was the least likely to have done so.
Therefore, $i^*$ must be the event to which earthquake $j$ is most
correlated. In general,
if $n_{ij}$ is  extremely small, then the correlation between $j$ and $i$ is
very strong, and {\it vice versa}. By this argument, the degree of
correlation between any two earthquakes $i$ and $j$ is
inversely proportional to $n_{ij}$.   Since the space-time-magnitude scales are
selected by the actual sequence of events, the variables $n_{ij}$ can be
considered to be self-organizing
tags of the underlying physical process governing seismicity.
 Note that singularities are
eliminated by taking a small scale cutoff in time (here $t_{\rm min}=
180$ sec) and a minimum spatial resolution (here $l_{\rm min} = 100$
meters). 

The metric defined by Eq.~\ref{eq:n} allows various classifications of
aftershocks. 
Therefore, the question of which is the better candidate to be the
foreshock of an event can be quantitatively decided.  Hierarchical
clusters of earthquakes emerge, in which the biggest event in the
cluster is called the main event, but where possibly later
aftershocks create their own sequences of aftershocks, whenever they
are able to ``steal'' aftershocks from the main event, and so on for
further generations of aftershocks.  Nevertheless, earthquakes are
automatically collected into hierarchically self-organized
clusters, without any special pre-analysis of single event properties.

In the language of modern complex network theory
\cite{albert,bornholdt02:_book}, 
what we achieve is a time-oriented growing
network where nodes (earthquakes) have internal variables (magnitude,
occurrence time, and location), and links between the nodes carry a
weight (the metric  $n_{ij}$) and are directed according
to the time orientation, from the older to the newer nodes.
Empirically, we find that both the distribution of outgoing links and
the cluster size distribution are scale free.  Due to the continuous
nature of the link variable, $n_{ij}$, no event is {\it a priori} purely an
aftershock or a main shock.  However, due to the broad distribution of
$n_{ij}$ observed, main shocks and aftershocks emerge as extreme limits
of a continuous spectrum of the extent to which any given event can be
considered to be a precursor or aftershock of other events in the sequence.

Our approach was inspired by a recent analysis of earthquake waiting
times by Bak et al.~\cite{bak02:_unified,corral03:_unified}.  They
introduced a space-time-magnitude scaling variable that allows a data
collapse of the distribution of waiting times between subsequent
earthquakes larger than a specified magnitude, occurring within grid cells
of a specified size, covering non-overlapping areas of the Earth.
Also, Abe and Suzuki found scale free networks for earthquakes in a
completely different context, where nodes representing these grid
cells were linked when subsequent earthquakes occurred in
them~\cite{abe_suzuki}.  However, neither of these works quantified
the correlation between an arbitrary pair of earthquakes, or dealt
with the subject of aftershock identification.

%%%%%%
\section{Data and parameters}\label{sec:data}
The catalog we have analyzed is maintained by the Southern California
Earthquake Data Center (it can be downloaded from the SCEDC web site
  http://www.scecdc.scec.org/ftp/catalogs/scsn), for which $\Delta m=0.1$. 
It is considered to be complete  for events with $m>2$. 
We use data ranging from January 1, 1984 to December 31, 2000. 
%In this period the three largest events were 
%Landers ($m=7.3$, June 28th, 1992),
%Hector Mine ($m=7.1$, Oct.~16th, 1999) and 
%Northridge ($m=6.7$, Jan.~17th, 1994).
In order to work with a well defined ensemble, a lower threshold on the
magnitude is introduced: events with magnitude smaller than 
$m_<$ are discarded.
For each event, its position $i$ in the sequence is  used as a label, and
we record the magnitude $m_i$, the occurrence time $T_i$
(measured in seconds from  midnight of the first day), 
and the  latitude and longitude of the epicenter 
(converted to angles measured in radians, $\theta_i$ and $\phi_i$ 
respectively).
The distance between two events $i$ and $j$ is then measured as the arc length
on the Earth's surface, $l_{ij}=R_0\arccos[ \sin(\theta_i)\sin(\theta_j) +
\cos(\theta_i)\cos(\theta_j)\cos(\phi_i-\phi_j) ]$, where 
the Earth radius is $R_0=6.3673 \times 10^6$ meters.

The $b$-value of the G-R law is $b \simeq 0.95$ for this data set,
while  $d_f\simeq 1.6$ was found by 
Corral~\cite{corral03:_unified} 
using a box counting procedure.  It is consistent with the correlation 
dimension we measure for most of our clusters.  
However, many of the statistical results we find
are not  sensitive to the precise value of $d_f$ or $b$.

With these units and values, the constant $C$ can be estimated using
Eq.~\ref{eq:n-mean}.  However, a precise evaluation of $C$ is not
possible, because $\bar n$ is the mean of a variable with huge
variations in space and time.  We have measured $\bar n$ for several
circular windows well inside the zone covered by the catalog, finding
$C\leq10^{-9}$.  For simplicity, our choice in this paper is
$C=10^{-9}$.  Most of our results are insensitive to the
precise value of $C$ because we focus on relative, rather than
absolute correlations between a pair of events.  Throughout this paper
we use, unless otherwise stated, the above mentioned values, and a
lower threshold $m_< = 2.5$.

To simplify notation, we denote the probability distribution of a
generic quantity $q$ as $P(q)$.  On finding distributions decaying as
power laws, a clearer result appears by binning the values of $P(q)$
in properly normalized bins of a width that grows geometrically with
$q$.

%%%%%%
\section{Method}\label{sec:method}

In the simplest implementation of the network, each new earthquake $j$ is
attached with a single link to the previous earthquake in the sequence
that minimizes $n_{ij}$, with a weight denoted as $n_{j}^*$.  Hence,
each link carries the extremal $n_{j}^*$ for the added node $j$ relative to
all previous nodes, and globally one obtains a growing directed tree.
Links with small $n_{j}^*$ indicate a stronger correlation between the
emitting node and the receiving one, and are expected to identify
events normally classified as aftershocks.  Weak links with large
$n_{j}^*$ arise when none of the previous events are sufficiently strong,
and close in space and time to event $j$.  
Thus, the strength of the link to an event $j$ is inversely
proportional to $n_j^*$.

A natural decomposition of the network
into clusters is achieved by then removing all weak links where 
$n_j^* > n_c$, and $n_c$ is a link threshold value.  The correlated events are
reliably detected when $n_c$ is less than one but not extremely small.  In the latter case,
correlated events detach, and a very fragmented network appears.
For large $n_c$ some uncorrelated events make links, and a
giant cluster appears.  The resulting space-time windows are concave
(see Figure~\ref{fig:window}, and Conclusions),  at variance
with the convex windows usually used.

%%%%%%
\section{Results}\label{sec:results}
\begin{figure}[!tb]
\includegraphics[height=85mm]{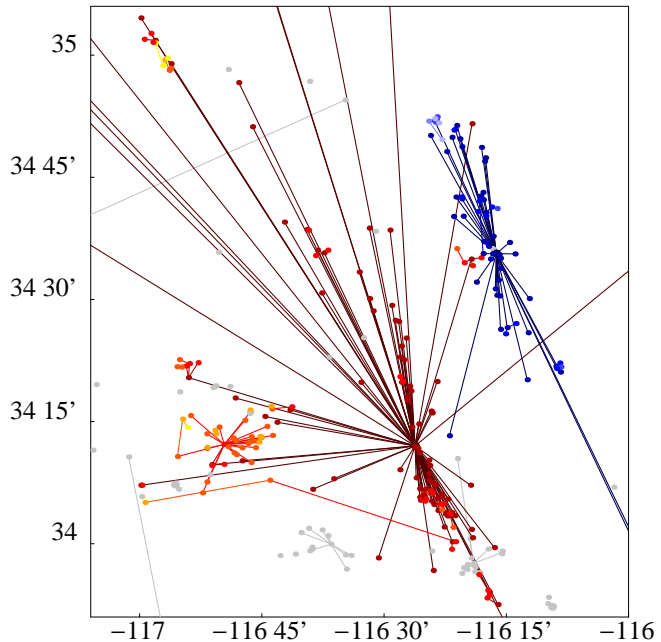}
\vskip 0.5truecm
\caption{Scale free earthquake network around Landers epicenter (red cluster) and
Hector Mine epicenter (blue cluster).  Colors fade with
the aftershock generation, from darker to lighter within each cluster.
Note that the big event following the Landers earthquake, giving rise to its own (orange) cluster
of aftershocks at $(-116.50', 34.10')$, is not a 
first generation aftershock, since it has no link from Landers. Here  $m_< = 4$, and $n_c=10^{-2}$.
\label{fig:net}}
\end{figure}

A part of the network constructed using this method is shown in 
Fig.~\ref{fig:net}.
Hierarchically organized clusters of earthquakes emerge, where the links 
join aftershocks with their most correlated predecessor.
%FIG %%%
%FIG %%%

\begin{figure}[!bt]
\includegraphics[width=65mm,angle=-90]{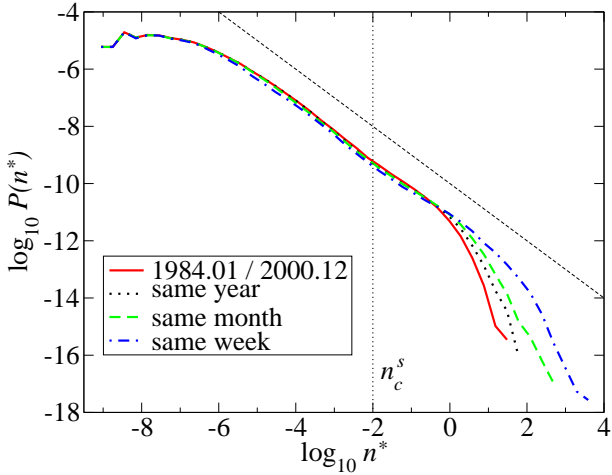}
\caption{The distribution of link weights, $n^*$, for sequences of
different temporal duration.  An average over all non-overlapping time
intervals of the same duration is shown.  The power law behavior is
stable to variations in the duration.  However, the cutoff moves to
smaller $n^*$ on increasing the measurement time interval as weakly
linked earthquakes find more correlated predecessors further in the
past.  The vertical dotted line represents the estimated transition
point, $n_c^s$, for the giant cluster. The straight dashed line has a
slope -1.
\label{fig:Px}}
\end{figure}

In order to quantitatively assess the properties of this network, we
start by analyzing the distribution  of link weights
$P(n^*)$.  This distribution exhibits power law
behavior with a slope $\simeq -1$ up to a cutoff, as shown in
Fig.~\ref{fig:Px}. Thus, the distribution of link strengths, $S=1/n^*$, is
also a power law, $P(S) \sim 1/S$. Such a broad, continuous
distribution, without particular characteristic peaks, indicates that
a division of earthquakes into rigid classes is intrinsically
impossible. Instead, a continuum of possibilities ranges from clear
aftershocks, which have an incoming link with  small $n^*$, to
events that are independent, with an incoming link of large $n^*$, but 
may emit many outgoing links with small $n^*$, and would be called main shocks.

%FIG %%%
%FIG %%% 

\begin{figure}[!tb]
\begin{center}
\includegraphics[width=45mm,angle=-90]{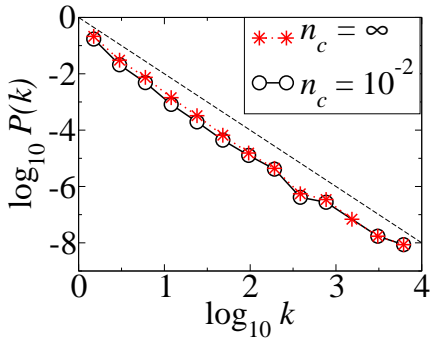}
\end{center}
\caption{The degree distribution of the network of earthquakes and
aftershocks.  The out-degree, $k$, is the number of aftershocks linked
to an earthquake.  The introduction of a threshold $n_c$ does not
alter the behavior.  The dashed line has slope $-2$, indicating a
scale free degree distribution, $P(k) \sim k^{-\gamma}$ with $\gamma
\approx 2$.
\label{fig:P_k}}
\end{figure}

The resulting network of earthquakes is scale free.  The number of
aftershocks of an earthquake is equal to the number, $k$, of outgoing
links from the node representing that event.  In the language of
network theory, this is called the out-degree of the node.
Fig.~\ref{fig:P_k} shows that earthquakes in  Southern California
form a scale free network, with an out-degree distribution
scaling over more than three decades, with an index $\gamma = 2.0(1)$.

Recently, many scale free networks with $P(k)\sim k^{-\gamma}$, have
been discovered~\cite{albert,bornholdt02:_book} in a broad variety of
contexts.  These include the Internet~\cite{faloutsos}, the
world-wide web~\cite{barabasiandalbert}, protein interaction and
genetic regulatory networks~\cite{wagner01,MaslovSneppen02}, and the
solar coronal magnetic field \cite{hughes_paczuski_et_al}. Both the
coronal field and the earthquake network share the property
that the strength of a link between a pair of nodes is also scale
free.

%FIG %%%
%FIG %%%

\begin{figure}[!tb]
\begin{center}
\includegraphics[width=50mm,angle=-90]{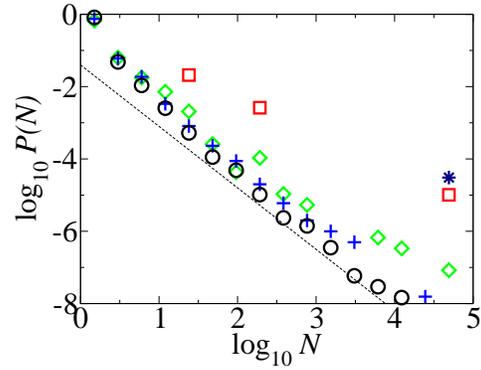}
\end{center}
\caption{Cluster size distribution for different link thresholds.
At large $n_c$, a giant cluster exists that is  well separated in size
from some remaining small ones.  Between $n_c=10^{-1}$ 
and $n_c=10^{-2}$, an apparently continuous
transition occurs where the finite  cluster distribution extends out toward the
giant cluster, and the distribution of cluster sizes exhibits
power law behavior.  The straight line has a slope -1.7. Symbols are
$(\ast\, n_c=10^2;\, \Box\, n_c=10;\, \Diamond\, n_c=1;\, {\bf +}\, 
n_c=10^{-1};\, \bigcirc\,  n_c=10^{-2})$.
\label{fig:P_cluster}}
\end{figure}

Lowering the link threshold $n_c$ from infinity, the fully connected
network breaks into clusters, in a percolation-like transition from a
giant cluster to a finite cluster regime.  We estimate the transition
to take place between link thresholds $n_c=10^{-1}$ and $n_c=10^{-2}$.  This estimate
is obtained by examining the distribution of cluster sizes, $N$, which
is the total number of earthquakes in a cluster, as a function of
$n_c$ (see Fig.~\ref{fig:P_cluster}).  Near the transition, the
cluster size distribution also appears to be scale free, $P(N)\sim
N^{-1.7(1)}$.  Furthermore, a scaling regime exists for a wide
range of link thresholds, indicating a relative insensitivity
to a  sharp separation between what are considered to be correlated and
uncorrelated events.  For clarity, we use the value $n_c^s=10^{-2}$ to
locate the transition point where the giant cluster emerges.  This
value is consistent with our ansatz, Eq.~\ref{eq:n}, which requires
that correlated events have $n$ values significantly less than one.  Network
clusters constructed with $n_c^s$ therefore only link strongly correlated events.
%FIG %%%
%FIG %%%

In Fig.~\ref{fig:Px}, we study the effect of changing the temporal
span of the catalog on the distribution of link weights.  The power
law behavior for strong links is stable, certainly up to $n_c^s$.
However, the cutoff in $P(n^*)$ for weak links decreases to smaller
$n^*$ values, when earthquakes can link to events at further distance
in the past.  For an ideal ``infinite'' catalog, we conjecture that
the cutoff value cannot be less than $n_c^s$. Indeed, below the
transition point a finite fraction of events stop having any
correlated predecessors.

\begin{figure}[!tb]
\includegraphics[width=65mm,angle=-90]{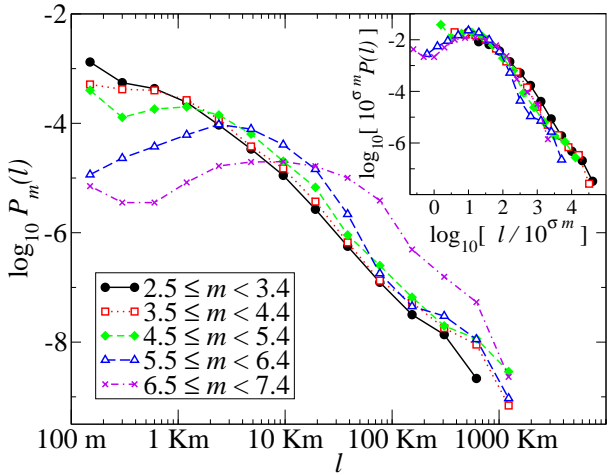}
\caption{ Link length distribution for different magnitudes of the
emitting earthquake, at $n_c^s$.  The length at maximum grows with
magnitude roughly as $l_{\rm max}\sim 10^{0.4 m}$, but the
distributions have a fat tail, extending up to hundreds of kilometers
even for intermediate magnitude events.  These distributions are
consistent with an hierarchical organization of events, where big
earthquakes preferentially link at long distance with intermediate
ones, which in turn link to more localized aftershocks, and so on.
Inset: Distributions rescaled according to Eq.~\ref{eq:Pl_resc} with
$\sigma=0.4$.
\label{fig:linkl}}
\end{figure}

We define the link length, $l$, as the distance between the epicenter
 of an aftershock and its linked predecessor.  The distribution of
 link lengths depends on the magnitude $m$ of the predecessor, being
 on average greater for larger $m$.  Dividing the link length
 distribution into classes depending on the magnitude of the
 predecessor, $P_m(l)$, a maximum in the distribution occurs, which
 shifts to larger $l$ on increasing $m$, as shown in
 Fig.~\ref{fig:linkl}.  This behavior is consistent with using larger
 space-time windows to collect aftershocks from larger events.

 However, the distribution of link lengths exhibits no cutoff at large
 distances, but rather decays slowly as a power law with $l$, up to
 the linear extent of the seismic region covered by the catalog.  The
 different distributions are consistent with a scaling ansatz: \beq
 P_m(l) \simeq 10^{-\sigma m} F \big( l/10^{\sigma m} \big) \,
\label{eq:Pl_resc}
\eeq where $l$ is measured in kilometers, $\sigma\approx 0.4$, and
$F(x)$ is a scaling function.  The tail of the scaling function is a
power law; i.e.  $F(x)\sim x^{-\lambda}$ with $\lambda \approx 2$ for
$x\gg 1$.  A data collapse using this ansatz is shown in the inset to
Fig.~\ref{fig:linkl}.  Such a slow decay at large distances calls into
question the use of sharply defined space windows for collecting
aftershocks, as already pointed out by Ogata~\cite{ogata98:_space}.

%FIG %%%
%FIG %%%

\begin{figure}[!tbp]
\includegraphics[width=65mm,angle=-90]{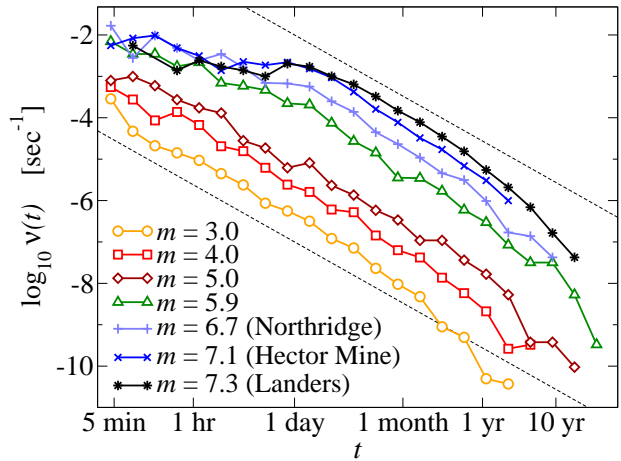}
\caption{The Omori law for aftershock rates. These rates are measured for 
aftershocks linked to earthquakes of different magnitudes, $m$,  using
$n_c^s$.  For each magnitude, the rate is consistent with the original
Omori law, Eq.~\ref{eq:Omori}, up to a cutoff time that depends on
$m$.  As guides to the eye, dashed lines represent a decay $\sim
1/t$.
\label{fig:Omori}}
\end{figure}

Figure~\ref{fig:Omori} shows the rate of aftershocks for the Landers,
Hector Mine, and Northridge events.  Aftershocks occurring at time $t$
after one of these events are binned into geometrically increasing
time intervals.  The number of aftershocks in each bin is then divided
by the temporal width of the bin to obtain a rate of earthquakes
per second.  The same procedure is applied to each remaining event,
not aftershocks of these three.  An average is made for the rate of
aftershocks linked to events having a magnitude within an interval $\Delta m$
of $m$.  Figure~\ref{fig:Omori} also shows the averaged results for $m=3$
(1710 events), $m=4$ (161 events), $m=5$ (28 events) and $m=5.9$ (4
events).

The collection of aftershocks linked to earthquakes of all magnitudes
is one of the main results of our method. Even intermediate magnitude
events can have aftershocks that persist up to years.  Earthquakes of
all magnitudes have aftershocks which decay according to the Omori
law~\cite{omori94b,utsu95:_omori}, \beq \nu(t) \sim
\frac{K}{c+t}\,,\label{eq:Omori} \quad {\rm for}\ t<t_{\rm cutoff}\eeq
where $c$ and $K$ are constant in time, but depend on the magnitude
$m$~\cite{utsu95:_omori,lise03} of the earthquake.  We find that the
Omori law persists up to a time $t_{\rm cutoff}$ that also depends on
$m$ as well as the link threshold, $n_c$.  Estimates of the cutoff
times for $n_c^s$ are $t_{\rm cutoff}\approx 3$ months for $m=3$,
and $t_{\rm cutoff} \approx 1$ year for $m=4$. For larger magnitudes,
it is difficult to distinguish $t_{\rm cutoff}$ from the temporal
duration of the data set.

%FIG %%%
%FIG %%%
The Omori law for aftershocks emerges as a result of our analysis,
although it is not part of the original ansatz, Eq.~\ref{eq:n}, used
to define aftershocks.  It has been extensively investigated over
decades, together with its modified version \cite{utsu95:_omori}
involving a scaling $\sim t^{-p}$.  The data shown in
Figure~\ref{fig:Omori} is consistent with the original Omori result,
$p=1$, for aftershocks of earthquakes of all magnitudes, once second
and further generations of aftershocks are excluded.  Our result is
also consistent with theoretical studies on stick-slip
motion~\cite{dieterich94:_frict,gomberg00:_coulomb}, which suggest $p
\approx 1$.

\section{Discussion}

Convex space time windows have been used since the 
1970's~\cite{gardner74:_after_window, keilis80:_after_window,knopoff82:_after_window,knopoff00:_after_window},
often with the size of the window determined by
the main shock magnitude. 
The performance of this procedure is  satisfactory for large 
earthquakes, although fixed window sizes may omit relevant
aftershocks.
Nevertheless, as a shortcoming,  
it can lead to distortions if many large
aftershocks occur. In this case, nothing can be said on the ``ownership''
of further aftershocks.

Different approaches to the problem of aftershocks collection were
proposed by several authors, sometimes with the aim to cure the former 
shortcomings.  For a review see \cite{molchan92:_after}.
Our method has some similarities with these approaches. 
For instance, Frohlich and Davis collected earthquakes in 
clusters~\cite{frohlich90:_after_SLC} by means of a different linking 
procedure. 
However, their analysis was done using a metric
of the form $\sim \sqrt{l^2 + {\rm const}\, t^2}$, 
which does not take into account the
magnitude of events, and has a space-time form at variance  with
measured earthquake correlations.
 
Maximum
likelihood methods~\cite{kagan91:_likel,ogata99:_review},
in the context of seismicity, usually 
start with an ansatz on the law governing 
aftershocks, typically the modified Omori law. It is further assumed that
seismicity is a non-stationary Poisson process.
Using a likelihood analysis with space, time and magnitude,
Ogata compared several forms of aftershocks distance 
distributions~\cite{ogata98:_space}, and showed  that an aftershock
rate of the form 
\beq
\nu_{m,l}(t) \sim \frac{10^{\alpha m}}{(c_l(m) + l)^\mu (c_t + t)^p}\,,
\label{eq:mod-Omori-magn-space}
\eeq was the most appropriate among his choices ($c_t, \alpha$, $p$
and $\mu$ are constant, while $c_l(m)$ is scaling with the magnitude
$m$ of the main shock).  Hence, he also concluded that fixed space
windows were not the best choice.  Indeed, our metric variable $n$ in
Eq.~\ref{eq:n} somewhat resembles his form of $\nu$.

However, our method is simpler to implement than likelihood methods.
 Furthermore, it does not
require an ansatz on the validity of the modified Omori law, or on the
type of statistical process that describes seismicity.  
Instead, the original Omori law is found as a result of our analysis.  
In addition, the physical argument leading to the variable
$n_{ij}$ also fixes the parameters in its definition, 
without the need to evaluate them by maximizing a likelihood.

One could object that the values  of $b$ and/or $d_f$ can depend on the
region of the Earth being considered, or may fluctuate depending on the
specific fault zone being studied.  However, the statistical
results we find, as shown in Figs.~3-7, are remarkably robust to variations
in either of these parameters.  Varying $d_f$ over a wide range, from $1$ to $3$
(using $d_f>2$ requires the introduction of event depths, see below)
does not alter considerably the distribution of outgoing links, which
retains its power law behavior with index $\gamma\approx 2$.  The
distribution of link weights, $n^*$, is even more
insensitive to variations of $b$ and $d_f$.
Also the Omori law with $p\approx 1$, shown in Fig.~7, does not depend
sensibly on the parameters, and holds for aftershocks linked to earthquakes of
all magnitudes.

The crust of the Earth has a finite width ($\approx 20$ Km in
California) in which events take place according to a
``three-dimensional'' fractal distribution, involving their depth.  It
is believed that there is a qualitative difference between small
earthquakes and large ones, the former producing ruptures smaller that
the crust width~\cite{scholz_book}.  Hence, our arguments may need to
be corrected at distances of the order of tens of kilometers.  We have
computed spatial distances through the three-dimensional Euclidean
metric distance, using an appropriately revised $d_f$ in
Eq.~\ref{eq:n}.  No significant departures from the results leading to
our present conclusions were found.

The introduction of more than one correlated predecessor for an event
will be the subject of a future investigation.  This is accomplished
by attaching links between all earthquake pairs where $n_{ij}< n_c$.
In this case, a general network, which is not tree-like emerges.  The
clustering of earthquakes could then be quantified in terms of the
clustering coefficient of the nodes in the
network~\cite{watts_strogatz98,MaslovSneppen02}.  In our view, an
earthquake network with nodes having multiple incoming links
represents a second order modeling of seismicity, the first being the
simple tree structure we have presented here.

\section{Conclusions}

We have introduced a metric to determine correlations between
earthquakes that takes into account known statistical properties
of seismicity.  By means of an appealingly simple yet quantifiable procedure, 
networks of earthquakes and aftershocks emerge, where the number
of aftershocks linked to any event is scale free with an index
$\gamma\approx2$.  The metric is constructed by looking backward in
time from any particular event and calculating an expected number of events that
would occur, compared to events that actually occurred.  If this
ratio is significantly less than one, then the preceding event is
correlated with the particular one.  This is reminiscent of Kierkegaard's
adage that life must be lived forward, but can only be understood
backward.

Due to the form of the metric $n$ measuring correlations, larger earthquakes collect 
aftershocks from larger space-time windows.  From Eq.~\ref{eq:n}, these windows have 
a spatial radius varying with
time as $r_{i}(T) = [n_c (T-T_i)^{-1}10^{b m_i}]^{1/d_f}$.
They span an hyperbolic space-time region (see Fig.~\ref{fig:window}), which is
at variance with the usual ``rectangular'' or convex windows, 
of constant radius up to a finite time.  In our method,
at early times after an earthquake, its aftershock collection window is
wider in space than it is at later times.

To our knowledge, the idea that an earthquake can be correlated to an
event very far away, if it occurs shortly after it, is new and
certainly unconventional.  But it is consistent with the hypothesis
that seismicity is a self-organized critical
phenomenon~\cite{bak_tang89,bak_book,turcotte99:_SOC}.  In that case,
some locations may be ``on the edge of giving an earthquake'' (or
toppling, according to the sandpile paradigm), and even a small
perturbation from an event far away could trigger them.  However, we
do not necessarily ascribe the correlations measured here to represent
a usual cause and effect relationship.  In the sandpile paradigm a
completely insignificant event, like adding one grain of sand to an
enormous pile, can trigger an arbitrarily large avalanche involving
the whole system.  Indeed, seismicity as one hierarchically correlated
self-organized critical process, generates the scale free network of
earthquakes and aftershocks.

Our results also suggest that modern network theory may be a useful
and illuminating way to approach the complexities of seismicity,
including perhaps problems related to prediction.  Our metric and
network construction may also have applications to other phenomena
with intermittent bursts such as, for instance, solar flares or even
turbulence.

%\bibliographystyle{./pnas}
%\bibliography{biblio_EQ}

\begin{thebibliography}{10}

\bibitem{turcotte_book}
Turcotte, D.~L.
\newblock (1997) {\em Fractals and Chaos in Geology and Geophysics}.
\newblock (Cambridge University Press, Cambridge), 2nd edition.

\bibitem{scholz_book}
Scholz, C.~H.
\newblock (2002) {\em The Mechanics of Earthquakes and Faulting}.
\newblock (Cambridge University Press, Cambridge), 2nd edition.

\bibitem{kagan}
Kagan, Y.~Y.
\newblock (1994) {\em Physica D} {\bf 77}, 160--192.

\bibitem{gutenberg-richter41}
Gutenberg, B. \& Richter, C.~F.
\newblock (1941) {\em Geol. Soc. Am. Bull.} {\bf 34}, 1--131.
\newblock Special papers.


\bibitem{omori94b}
Omori, F.
\newblock (1894) {\em J. Coll. Sci. Imp. Univ. Tokyo} {\bf 7}, 111--200.

\bibitem{bak02:_unified}
Bak, P., Christensen, K., Danon., L.,  \& Scanlon, T.
\newblock (2002) {\em Phys.\ Rev.\ Lett.} {\bf 88}, 178501.

\bibitem{gardner74:_after_window}
Gardner, J. \& Knopoff, L.
\newblock (1974) {\em Bull.\ Seism.\ Soc.\ Am.} {\bf 64}, 1363--1367.

\bibitem{keilis80:_after_window}
Keilis-Borok, V., Knopoff, L.,  \& Rotwain, I.
\newblock (1980) {\em Nature} {\bf 283}, 259--263.

\bibitem{knopoff82:_after_window}
Knopoff, L., Kagan, Y.,  \& Knopoff, R.
\newblock (1982) {\em Bull.\ Seism.\ Soc.\ Am.} {\bf 72}, 1663--1675.

\bibitem{knopoff00:_after_window}
Knopoff, L.
\newblock (2000) {\em Proc.\ Natl.\ Acad.\ Sci. USA} {\bf 97}, 11880--11884.

\bibitem{hirata89:_d_fract}
Hirata, T.
\newblock (1989) {\em J.\ Geophys.\ Res.} {\bf 64}, 7507--7514.

\bibitem{albert}
Albert, R. \& Barab\'asi, A.L.
\newblock (2002) {\em Rev.\ Mod.\ Phys.} {\bf 74}, 47--97.

\bibitem{bornholdt02:_book}
Bornholdt, S. \& Schuster, H.~G., eds.
\newblock (2002) {\em Handbook of Graphs and Networks}.
\newblock (Wiley-VCH).

\bibitem{corral03:_unified}
Corral, A.
\newblock (2003) {\em Phys.\ Rev.\ E} {\bf 68}, 035102(R).

\bibitem{abe_suzuki}
Abe, S. \& Suzuki, N.
\newblock (2002) Scalefree network of earthquakes.
\newblock preprint cond-mat/0210289.

\bibitem{faloutsos}
Faloutsos, M., Faloutsos, P.,  \& Faloutsos, C.
\newblock (1999) {\em Proc.\ ACM SIGCOMM, Comput.\ Commun.\ Rev.} {\bf 29},
  251-262.

\bibitem{barabasiandalbert}
Barab\'asi, A.L. \& Albert, R.
\newblock (1999) {\em Science} {\bf 286}, 509--512.

\bibitem{wagner01}
Wagner, A.
\newblock (2001) {\em Mol. Biol. Evol.} {\bf 18}, 1283.

\bibitem{MaslovSneppen02}
Maslov, S. \& Sneppen, K.
\newblock (2002) {\em Science} {\bf 296}, 910.

\bibitem{hughes_paczuski_et_al}
Hughes, D., Paczuski, M., Dendy, R.~O., Helander, P.,  \& McClements, K.~G.
\newblock (2003) {\em Phys.\ Rev.\ Lett.} {\bf 90}, 131101.

\bibitem{ogata98:_space}
Ogata, Y.
\newblock (1998) {\em Ann.\ Inst.\ Statist.\ Math.} {\bf 50}, 379--402.

\bibitem{utsu95:_omori}
Utsu, T., Ogata, Y.,  \& Matsu'ura, R.~S.
\newblock (1995) {\em J.\ Phys.\ Earth} {\bf 43}, 1--33.

\bibitem{lise03}
Lise, S., Paczuski, M.,  \& Stella, A.~L.
\newblock (2003) Scaling law for earthquake hazard after a main shock.
\newblock Submitted.


\bibitem{dieterich94:_frict}
Dieterich, J.
\newblock (1994) {\em J.\ Geophys.\ Res.} {\bf 99}, 2601--2618.

\bibitem{gomberg00:_coulomb}
Gomberg, J., Beeler, N.,  \& Blanpied, M.
\newblock (2000) {\em J.\ Geophys.\ Res.} {\bf 105}, 7857--7871.

\bibitem{molchan92:_after}
Molchan, G.~M. \& Dmitrieva, O.~E.
\newblock (1992) {\em Geophys.\ J.\ Int.} {\bf 192}, 501--516.

\bibitem{frohlich90:_after_SLC}
Frohlich, C. \& Davis, S.~D.
\newblock (1990) {\em Geophys.\ J.\ Int.} {\bf 100}, 19--32.

\bibitem{kagan91:_likel}
Kagan, Y.~Y.
\newblock (1991) {\em Geophys.\ J.\ Int.} {\bf 106}, 135--148.

\bibitem{ogata99:_review}
Ogata, Y.
\newblock (1999) {\em Pure Appl.\ Geophys.} {\bf 155}, 471--507.

\bibitem{watts_strogatz98}
Watts, D.~J. \& Strogatz, S.~H.
\newblock (1998) {\em Nature} {\bf 393}, 440--442.

\bibitem{bak_tang89}
Bak, P. \& Tang, C.
\newblock (1989) {\em J.\ Geophys.\ Res. -- Solid Earth} {\bf 94}, 15635-15637.

\bibitem{bak_book}
Bak, P.
\newblock (1996) {\em How Nature Works: The Science of Self-Organized
  Criticality}.
\newblock (Copernicus, New York).

\bibitem{turcotte99:_SOC}
Turcotte, D.~L.
\newblock (1999) {\em Rep.\ Prog.\ Phys.} {\bf 62}, 1377--1429.


\end{thebibliography}

\end{document}